\documentclass[12pt]{iopart}

\usepackage{iopams}  
\usepackage{graphicx}
\usepackage[usenames,dvipsnames]{color}
\usepackage{units}
\usepackage{subfigure}
\usepackage[pdfpagelabels,bookmarks=true,colorlinks,linkcolor=black,urlcolor=black,citecolor=black]{hyperref}
\usepackage{memhfixc}

\newcommand{\KE}{KATRIN experiment}

\newcommand{\Mac}{MAC-E filter}

\newcommand{\ie}{\emph{i.e.}}
\newcommand{\eg}{\emph{e.g.}}
\newcommand{\bsym}{\boldsymbol} 		

\begin{document}
\title{A UV LED-based fast-pulsed photoelectron source for time-of-flight studies}
\author{K.~Valerius$^{1,}$\footnote{corresponding author}, M.~Beck$^1$, H.~Arlinghaus$^1$,
  J.~Bonn$^2$, V.~M.~Hannen$^1$, H.~Hein$^1$, B.~Ostrick$^{1, 2}$, S.~Streubel$^1$, Ch.~Weinheimer$^1$, and M.~Zbo\v{r}il$^{1, 3}$}
\address{$^1$ Institut f\"ur Kernphysik, Westf\"alische Wilhelms-Universit\"at
  M\"unster, Germany}
\address{$^2$ Institut f\"ur Physik, Johannes Gutenberg-Universit\"at Mainz,
  Germany}
\address{$^3$ Nuclear Physics Institute ASCR, \v{R}e\v{z} near Prague, Czech Republic}
\ead{valerius@uni-muenster.de}
%
%
\begin{abstract}
We report on spectroscopy and time-of-flight measurements using an $\unit[18]{keV}$ fast-pulsed photoelectron source of adjustable intensity, ranging from single photoelectrons per pulse to 5 photoelectrons per $\mu$s at pulse repetition rates of up to 10 kHz. Short pulses between $\unit[40]{ns}$ and $\unit[40]{\mu s}$ in length were produced by switching light emitting diodes with central output wavelengths of $\unit[265]{nm}$ and $\unit[257]{nm}$, in the deep ultraviolet (or UV-C) regime, at kHz frequencies. Such photoelectron sources can be useful calibration devices for testing the properties of high-resolution electrostatic spectrometers, like the ones used in current neutrino mass searches.
\end{abstract}
%
%
\maketitle
\section{Introduction}
\label{sec:introduction}
We describe the application of ultraviolet light emitting diodes to produce
photoelectrons for a calibration source at electron energies in the keV range.

Electrostatic filters with magnetic adiabatic collimation ({\lq\Mac s\rq}) \cite{hsu-hirshfield,beamson,kruit-read,picard-nimb,lobashev85} have
prov\-en to be ideal instruments for neutrino mass searches based on
high-precision measurements of the tritium beta decay spectrum near the
endpoint \cite{kraus,lobashev03,kdr,otten-weinh}. The upcoming Karlsruhe Tritium Neutrino experiment (KATRIN) \cite{kdr}  aims at increasing the sensitivity on $m(\nu_\mathrm{e})$ by an order of magnitude with respect to its predecessors. This requires an electron spectrometer with a
resolving power of $E/\Delta E = 2\cdot 10^4$, corresponding to an energy
resolution of $\Delta E \lesssim \unit[1]{eV}$ at the endpoint of the tritium beta spectrum ($E =
\unit[18.6]{keV}$). 

Calibration sources at these keV energies offering line
widths of the order of or even below this resolution are necessary for
different purposes:

 For continuous monitoring of fluctuations of the retardation potential, an
 electron source with high stability both of the energy and the intensity is
 given by a nuclear/atomic standard using $\unit[17.8]{keV}$ conversion
 electrons from $^{83\mathrm{m}}$Kr \cite{picard-zphys}. The K-shell
 conversion line exhibits a natural width of $\Gamma = \unit[2.7]{eV}$ \cite{campbell}, which
 is small enough to monitor the retardation potential, but too broad for 
 detailed studies of the transmission properties of the {\Mac}. Such a
 conversion electron 
 source is presently being developed for KATRIN \cite{kdr,diss_bea}.

However, in order to investigate the details of the  
transmission properties of the {\Mac}, an
electron source in the keV range is needed, which can deliver fast-pulsed
electrons with a sub-eV and well-defined energy spread 
and tunable beam intensities
between single electrons and up to $10^5$ electrons per second.

A pulsed electron source can serve as a means to test a particular
``time-of-flight'' operational mode of the {\Mac} that turns it from a
high-pass into a band-pass filter and at the same time allows a more
detailed and faster characterization of its transmission properties. This idea
has been introduced in \cite{bonn}, where a first experimental study
validating the concept is described. Such a source requires a fast timing with
pulse lengths $\tau$ both smaller than the time-of-flight of the electrons through
the {\Mac} and smaller than the expected variation of the
time-of-flight\footnote{In addition, a fast data acquisition and analysis is
  required, with a timing resolution again smaller than both the
  time-of-flight and its expected variation.}.

\begin{figure*}[!tb]
 \centering
\includegraphics[width=0.72\textwidth]{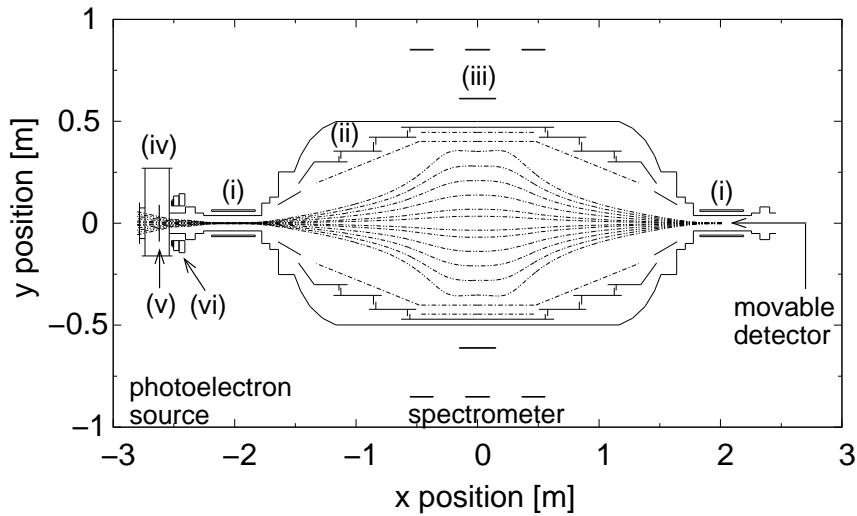}
\caption{Schematic of the experimental setup to test the photoelectron
  source. From left to right: vacuum chamber with photoelectron source,
  electrostatic filter with magnetic adiabatic guiding and detector. The
  details shown in the sketch include: (i) two superconducting solenoids to
  produce the magnetic guiding field of the {\Mac} (ii) the electrode
  configuration comprising a vacuum tank on ground potential and an inner
  high-voltage electrode system, (iii) field-shaping air coils, (iv) the
  vacuum chamber of the photoelectron source
  (see figure \ref{fig:ellitopf-details}), (v) the stainless steel cathode and
  (vi) an additional water-cooled coil for local enhancement of the magnetic
  field strength. Magnetic field lines connecting the photocathode and the
  detector are indicated as dashed curves. The analyzing plane of the
  spectrometer at $x = 0$ is defined by the maximum of the retardation
  potential $|U_\mathrm{spec}|$ coinciding with the minimum magnetic field strength
  $B_\mathrm{min}$.\label{fig:experimental-setup}}
\end{figure*}
\begin{figure*}[tb]
 \centering
\includegraphics[width=0.62\textwidth]{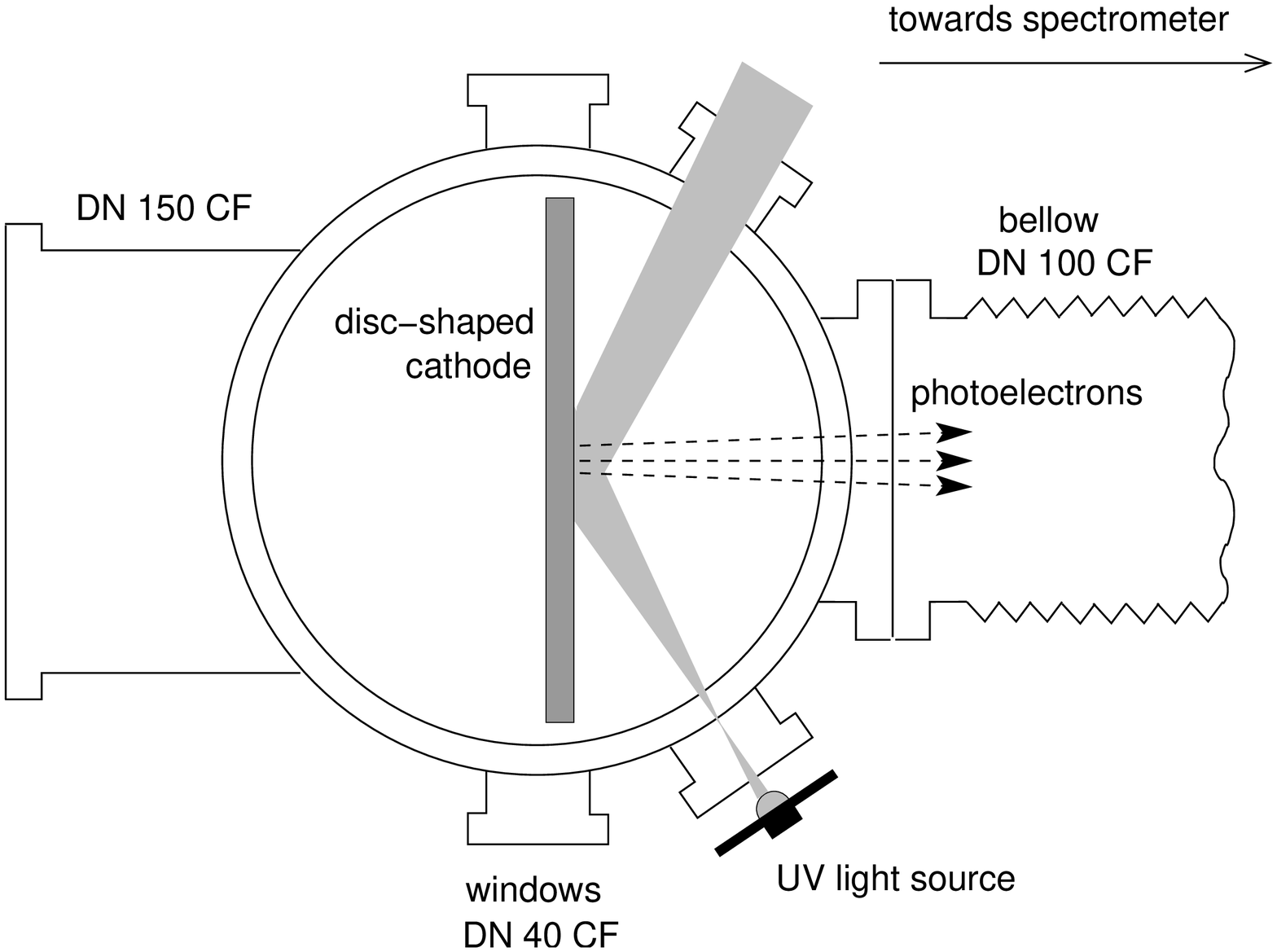}
\caption{Schematic of the photoelectron source: top view of the vacuum chamber
  housing the disc-shaped cathode on high voltage ($U_\mathrm{source}$) seen edge-on. The diameter of the stainless steel disc is $D = \unit[180]{mm}$ and its thickness about $d = \unit[4]{mm}$ (with a $\varnothing \unit[12]{mm}$ rounded bead at the rim to prevent field emission when high voltage is applied). The chamber is equipped with four windows, two of which are made of UV-light transmitting quartz glass that allows to illuminate the cathode with an external UV light source. Photoelectrons are accelerated in the strong electric field and later guided magnetically into the spectrometer. (The  magnetic field lines are not shown.)\label{fig:ellitopf-details}}
\end{figure*}

 A potentially suitable method to set up an electron source fulfilling the
 abovementioned requirements is to irradiate a cathode on high voltage with
 narrow-band UV light. This can be realized in a simple manner using modern
 deep-UV light emitting diodes (LEDs).
Typical work functions $\Phi$ of metals lie in the range of about $\unit[4.2]{eV}$
(Ag) to $\unit[5.1]{eV}$ (Au), with $\Phi = \unit[4.4(2)]{eV}$ \cite{picard-nimb} for stainless
steel, corresponding to a wavelength of the light of $\lambda = \frac{hc}{E}
\approx \unit[282]{nm}$. Standard laser technology can
also be used to obtain output wavelengths in the deep-UV range. However, such (high-power) laser systems are rather expensive, immobile and require particular safety installations in the laboratory. By contrast, light emitting diodes operating in the deep-UV domain are advantageous due to their versatility and ease of use, albeit offering only a fraction of the optical output power\footnote{At emission wavelengths as short as $\unit[247]{nm}$, an optical power output of $\unit[300]{\mu W}$ for continuous operation and $\lesssim \unit[10]{mW}$ in pulsed mode has been reported \cite{deng}.} as well as inferior spectral and beam profile characteristics compared to laser systems. Certain fields of application, however, benefit from this feature of moderate and highly controllable output power combined with fast pulsing (pulse length $\tau \gtrsim \unit[25]{ns}$, switching slopes of the order of a few ns, see for example references \cite{gaska-zhang-2005,shatalov-zhang-gaska-2003}) at repetition rates of $\unit[(1 - 10)]{kHz}$. In the particular application of photoelectron creation, the drawback of line half-widths of the order of $\delta \lambda \approx \unit[15]{nm}$ typically obtained for such LEDs can partially be compensated by an appropriate matching of the work function $\Phi$ of the photocathode material and the photon energy $E_\mathrm{photon}$. If $\Phi$ and $E_\mathrm{photon}$ are chosen such that only photons from the high-energy part of the distribution can cause photoemission, the energy spread of the photoelectrons will be reduced accordingly.

In this article we will show that photoelectrons created with UV light from a pulsed
LED can be used to investigate the transmission function of electrostatic
electron spectrometers and to perform time-of-flight measurements. The article is
structured as follows: Section \ref{sec:setup} contains an overview of the
setup and equipment used for our measurements, focusing on the photoelectron
source and the retardation spectrometer. In section \ref{sec:photoelectrons}
we demonstrate the use of short UV light pulses to obtain tunable
photoelectron rates. 
To determine the angular emission and the energy width 
of the photoelectron source, we measured with the MAC-E-Filter at Mainz
an integrated energy spectrum and a time-of-flight
spectrum, which are both presented in section
\ref{sec:tof-measurement}. The paper concludes with a discussion of the
results and an outlook in section \ref{sec:outlook}.

\section{Experimental setup}
\label{sec:setup}
Figure \ref{fig:experimental-setup} shows the setup used for the measurements reported on in this paper, the three main components being the photoelectron source, the electrostatic spectrometer and an electron detector. 

\begin{figure}[htb]
 \centering
\includegraphics[height=0.5\textwidth,angle=-90]{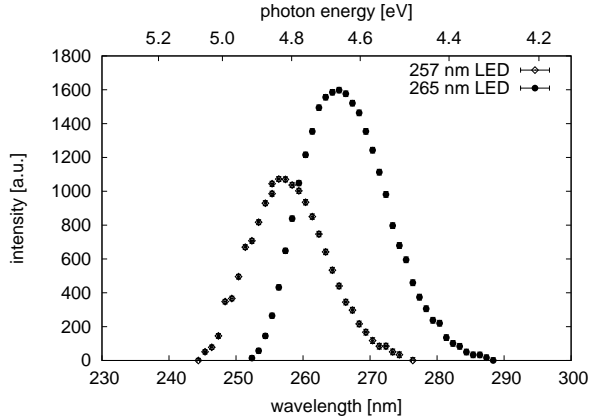}
\caption{Spectra of deep-UV LED type T9B25C ({\boldmath$\diamond$}) and T9B26C
  ($\bullet$), recorded with a grating spectrograph and a silicon PIN
  diode. In addition to the ultraviolet peak, a less intense emission
  component of visible light is present (not shown). This measurement was made at $I_\mathrm{LED}=\unit[12]{mA}$ (continuous operation of the LED).\label{fig:led-spectra}}
\end{figure}

\paragraph{Photoelectron source:}
~A sketch of the photoelectron source is presented in
figure~\ref{fig:ellitopf-details}. The photocathode consists of a mechanically
polished plane disc made of stainless steel with a diameter of $D = \unit[180]{mm}$ and a thickness of $d = \unit[4]{mm}$, encased in a cylindrical vacuum chamber with an inner diameter of $\unit[200]{mm}$. Stainless steel was selected as the cathode material because of its easy hand\-ling and ready machinability, as well as its work function of $\Phi = \unit[4.4(2)]{eV}$, which is just within the photon energy reach of commercially available UV LEDs. As light sources we employed UV LEDs of the type T9B26C and T9B25C from Seoul
Semiconductor Co., Ltd, with central wavelengths of $\lambda_\mathrm{central} = \unit[265]{nm}$ and $\unit[257]{nm}$ at a spectral width of $\Delta \lambda_\mathrm{FWHM} = \unit[15.3]{nm}$ and $\unit[13.8]{nm}$, respectively (compare measured light spectra in figure~\ref{fig:led-spectra}). The
LEDs provide a maximum cw optical power output of
$P_\mathrm{opt} \leq \unit[400]{\mu W}$ (T9B26C) and $P_\mathrm{opt} \leq
\unit[150]{\mu W}$ (T9B25C) \cite{seoul-led-265,seoul-led-255}, and each
is equipped with a focusing ball lens. The energy spread derived from the width of the
UV light peak is $\mathrm{FWHM} = \unit[0.27]{eV}$ for the 265 nm LED at a central
energy of $\unit[4.68]{eV}$ and $\mathrm{FWHM} = \unit[0.26]{eV}$ for the 257 nm LED at a
central energy of $\unit[4.82]{eV}$. By comparing this with the work function of
stainless steel of $\unit[4.4(2)]{eV}$ it is apparent that the low-energy
part of the UV light may be cut off by the work function.
However, the photocathode surface was not especially prepared 
and the vacuum at the photocathode of $p \lesssim \unit[5 \cdot 10^{-9}]{mbar}$ was not good enough to guarantee
the homogeneity of the work function of the photocathode. In section \ref{sec:tof-measurement} we
determine the energy width of the photoelectrons to be compatible with a Gaussian standard deviation of about 0.2~eV, which matches the numbers
of the expected work function inhomogeneity of 0.2~eV and energy width of the incident photons.

\begin{figure}[htb]
 \centering
\includegraphics[width=0.3\textwidth]{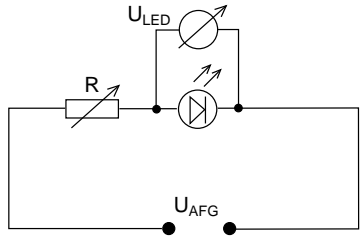}
\caption{Circuit for driving the UV LED via the function generator (AFG). The
  voltage $U_\mathrm{AFG}$ is supplied to the UV LED in series with a resistor
  of $R=\unit[90]{\Omega}$. The voltage at the UV LED
  $U_\mathrm{LED}$ was typically $\unit[5]{V}$ to
  $\unit[7.6]{V}$. The corresponding current was measured via the voltage drop
  across the resistor $R$ and varied between $I_\mathrm{LED}<\unit[1]{mA}$ and
  $I_\mathrm{LED} \approx \unit[20]{mA}$ in pulsed operation of the LED.\label{fig:uvcircuit}}
\end{figure}

The UV LED was supplied with pulses of varying amplitude and width $\tau$, with
rise/fall times of $\leq\unit[5]{ns}$, from a function generator Tektronix
AFG 3102 (figure~\ref{fig:uvcircuit}). The light intensity was controlled by the
voltage $U_\mathrm{LED}$. The number of photoelectrons that were produced per pulse
was determined by the interplay of $U_\mathrm{LED}$ and $\tau$. The UV LED was
placed in front of a UV-transparent window outside the vacuum chamber. The
resulting large distance of $\approx \unit[17]{cm}$ between the LED position
and the photocathode caused a widening of the UV light beam and a rather large
spot size of area $\approx \unit[2]{cm^2}$ on the cathode, as the spacing was
not matched to the focal length of the ball lens (see
figure~\ref{fig:ellitopf-details}). A residual gas pressure of $p \lesssim
\unit[5 \cdot 10^{-9}]{mbar}$ inside the vacuum chamber containing the
photoelectron source was maintained by means of a turbomolecular pump. The
combination of the spectrometer solenoids (see next paragraph) and an
additional water-cooled copper coil allowed us to vary the magnetic field
strength $B_\mathrm{source}$ at the location of the photocathode between $\sim
\unit[0.02]{T}$ and $\sim \unit[0.03]{T}$.

\paragraph{High-resolution spectrometer and electron detector:}~\\ The electron
spectrometer of the Mainz neutrino mass experiment \cite{picard-nimb} was
used for our measurements. This spectrometer of {\Mac} type consists of an
electrostatic retardation filter combined with an inhomogeneous magnetic
guiding field. Figure~\ref{fig:experimental-setup} shows the two
superconducting solenoids, which produce a strong magnetic field $B_\mathrm{max}$
that is attenuated towards the center of the spectrometer ($x = 0$ in figure~\ref{fig:experimental-setup}) by about 4 orders of
magnitude to a minimum value $B_\mathrm{min}$ (an air coil system helps to increase the field homogeneity). This results in an expansion of the magnetic flux tube, visualized in figure~\ref{fig:experimental-setup} by a widening of the magnetic field lines. Electrons starting from a source in a higher magnetic field will spiral around the magnetic field lines. As the magnetic field strength continuously decreases towards the center of the spectrometer, the cyclotron component $E_\perp$ of the kinetic energy will be reduced and transformed into longitudinal kinetic energy $E_\parallel$ according to the relation 
\begin{equation} 
 \mu =  \frac{E_\perp}{B} = \mathrm{const.}
\label{equ:magn-moment}
\end{equation}
stating that, in the adiabatic limit\footnote{If the relative change of the
  magnetic field strength is small in comparison to the cyclotron frequency
  $\omega_\mathrm{c}$, \ie, $|
  \frac{1}{B}\frac{\mathrm{d}\bsym{B}}{\mathrm{d}t} | \ll \omega_\mathrm{c}$,
  the conversion between $E_\perp$ and $E_\parallel$ is said to take place
  adiabatically. The concept of adiabaticity is discussed for example in
  references \cite{kruit-read,otten-weinh}.}, the orbital magnetic moment
$\mu$ of the electron is a conserved quantity\footnote{The adiabatic
invariant is given here in the non-relativistic limit.}.

We see that the energy component $E_\parallel$, which is analyzed by an
electrostatic retardation potential $U_\mathrm{spec}$, is maximized when the
magnetic field reaches its minimal value $B_\mathrm{min}$ in the so-called
``analyzing plane'' ($x = 0$ in figure~\ref{fig:experimental-setup}). Only those
electrons with sufficient axial kinetic energy, $E_\parallel >
qU_\mathrm{spec}$, with $q=-e$ being the electron charge, will be able to pass the filter and get re-accelerated towards the detector at the exit of the spectrometer.
The energy resolution $\Delta E$ of the {\Mac} thus follows directly from
equation~(\ref{equ:magn-moment}) by assuming that an electron starts at
$B_\mathrm{max}$ with a kinetic energy $E_\mathrm{kin}$ that resides entirely
in the cyclotron component. In this case, $\Delta E$ corresponds to the amount
of transverse energy $(E_\perp)_\mathrm{A,\,max}$ that is left at the
electrostatic analyzing plane of the filter after the adiabatic transformation
according to equation~(\ref{equ:magn-moment}):
\begin{equation}
 \Delta E = (E_\perp)_\mathrm{A,\,max} = E_\mathrm{kin} \cdot \frac{B_\mathrm{min}}{B_\mathrm{max}}.
\label{equ:energyres}
\end{equation}
The electrode system comprises a vacuum tank on ground potential, a set of several cylindrical retardation
electrodes on negative high voltage, and a grid electrode made of wires
\cite{flatt-paper}. In our measurements, the magnetic field inside the
superconducting solenoids was set to $B_\mathrm{max} = \unit[6]{T}$, while the
field strength $B_\mathrm{min}$ in the analyzing plane of the spectrometer was tuned to values
between $\unit[0.3]{mT}$  and $\unit[0.5]{mT}$ using the air coil
system. Inserting the values into equation~(\ref{equ:energyres}), these settings
result in an energy resolution  for $\unit[18]{keV}$ electrons of $\Delta E 
\approx \unit[1]{eV}$ or $\approx \unit[1.5]{eV}$, respectively. 

Since the energy selection is performed by the electrostatic filter, the
electron detector merely serves as a counter to measure an integrated spectrum
of electrons overcoming the retardation potential at a specific filter
threshold $qU_\mathrm{spec}$. We chose a windowless Si-PIN diode (type Hamamatsu S3590-06)
of size $\unit[9 \times 9]{mm}^2$ as electron detector. The magnetic field at the
location of the detector was $B_\mathrm{det}=\unit[0.34]{T}$. Since the
magnetic field at the plate of the photocathode was $\unit[0.02-0.03]{T}$, the
conservation of the magnetic flux requires that an area of $\unit[9-14]{cm}^2$
of the plate was imaged onto the detector. This is significantly larger than
the area of $\unit[2]{cm^2}$ which is illuminated by the UV LED. Therefore it
can be assumed that all photoelectrons will be imaged onto the detector.
The detection efficiency is close to 100~\% for the following reasons:
a) The photoelectrons impinge on the Si-PIN diode with an energy of 18~keV
giving a charge signal well above the noise level, b) the dead layer of 
this Si-PIN diode is so small\footnote{For a predecessor of this
Si-PIN diode the dead layer has been determined to amount to $45~\mu$g/cm$^2$
\cite{wei92}.} that the electrons only lose about 0.5~keV there, c)
photoelectrons backscattered in the detector will be reflected back to the 
detector by the magnetic mirror and
the electric potential barrier of the MAC-E filter and thus eventually deposit all of their energy in the detector.
\paragraph{Data acquisition and analysis methods:} 
~We recorded the signal of the preamplifier of the Si-PIN diode with a Flash-ADC system (Struck
SIS3301-105 FADC, 14~bit, 100~MHz, 10~ns sampling width), which was triggered
by the UV LED control pulse from the function generator. 
These digitized signals were first digitally filtered using a Butterworth
bandpass filter with lower and upper filter cutoff frequencies of $\unit[1]{kHz}$
and $\unit[1]{MHz}$, respectively, and subsequently
differentiated. The differential filtered signals were then analyzed for
electron events. Events with a time separation smaller
than $\unit[755]{ns}$ were counted as coincident events. For single
electron events the time resolution was $\Delta t_\mathrm{DAQ,\, analysis} \approx \unit[60]{ns}$\footnote{This time resolution was mainly limited 
by the fact that the energy deposit of the $\unit[18]{keV}$ electrons 
in the detector was not much more than the electronic noise level 
of the detector, which required the high-pass filtering with an edge 
frequency of $\unit[1]{MHz}$.}.
\section{Generation of few or single photoelectrons from short UV pulses}
\label{sec:photoelectrons}
In order to check the usability of these photoelectrons for a characterization
of the properties of {\Mac}s, the statistics of photoelectron creation was
investigated. Main questions are 
\begin{itemize}
 \item[--] how many photoelectrons are created, and how is their number influenced by the operating parameters of the UV LED,
 \item[--] what is the multiplicity distribution of the detected photoelectrons, and especially 
 \item[--] what is the total yield of single electrons in one pulse and
   what is their relative fraction.
\end{itemize}

\begin{figure}[!thb]
 \centering
\includegraphics[width=0.49\textwidth]{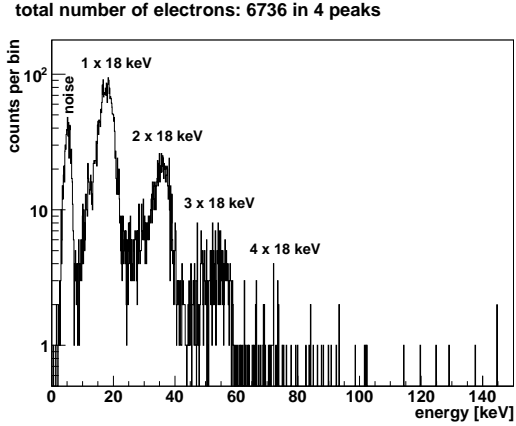}
\caption{Photoelectron spectrum measured with the Si-PIN diode (surplus energy  $E_\mathrm{surplus} = q(U_\mathrm{source} -
  U_\mathrm{spec}) = \unit[18.0]{keV}  - \unit[17.943]{keV} =
  \unit[0.057]{keV}$). Applied LED voltage $U_\mathrm{LED} = \unit[7.6]{V}$,
  pulse width $\tau = \unit[200]{ns}$, and repetition rate
  $\unit[1]{kHz}$. Electrons are detected at the energy
  $E_\mathrm{photoel.}^\mathrm{single} = qU_\mathrm{source}$ corresponding to
  the potential supplied to the pho\-to\-ca\-thode, and at various multiples
  of this energy, $E^n_\mathrm{photoel.} = n \cdot qU_\mathrm{source}$. This
  is due to the fact that at high electron flux the time resolution is not
  sufficient to resolve single electrons arriving quasi-simultaneously. For the
  calibration the energy
  loss in the deadlayer of the Si-PIN diode of about $\unit[0.5]{keV}$ was neglected. \label{fig:multiplepeaks}}
\end{figure}

\begin{figure}[!htb]
\centering
\subfigure[$U_\mathrm{LED}$ varied, $\tau = 40\mu \mathrm{s}$]{\includegraphics[width=0.32\textwidth,angle=-90]{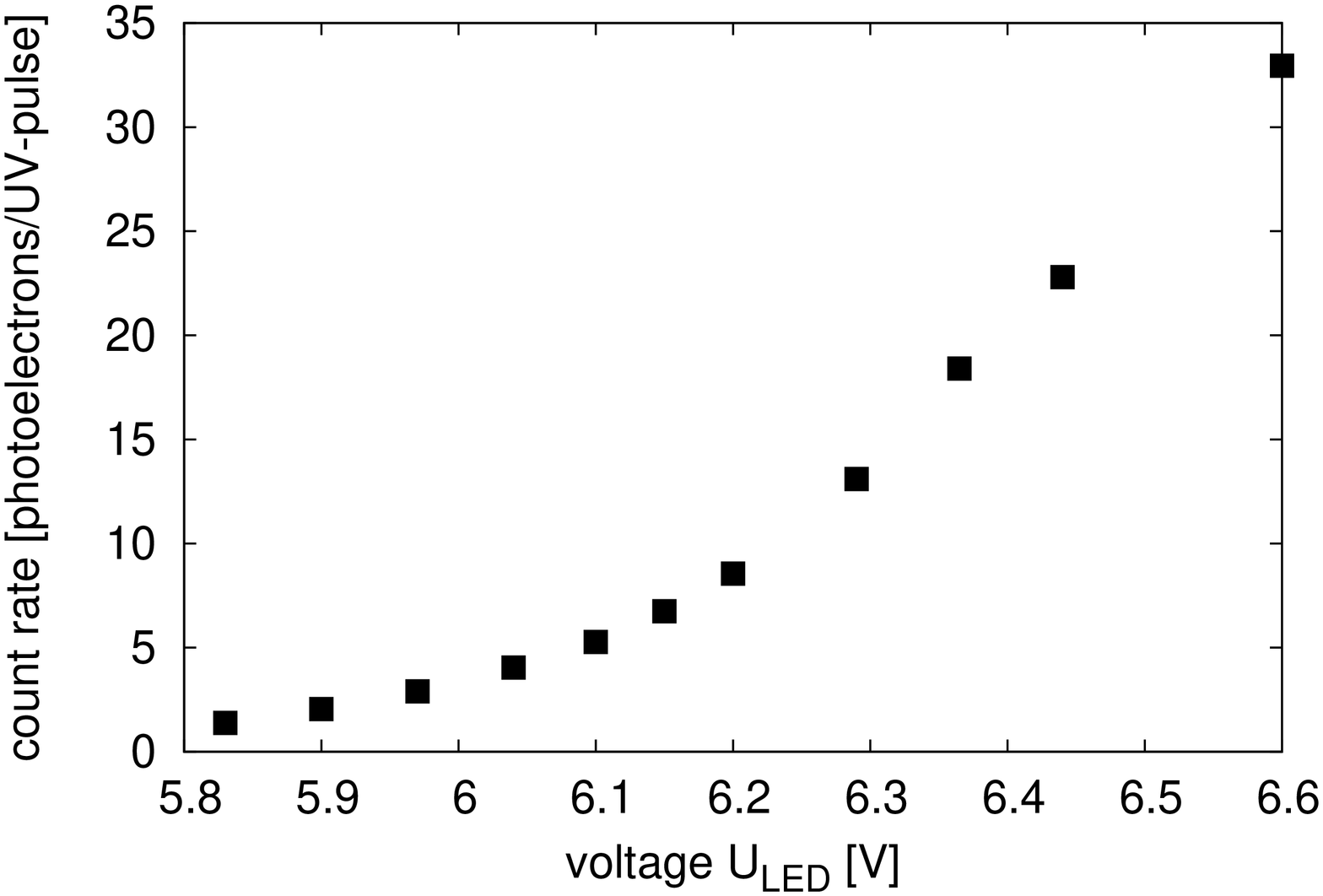}\label{fig:ledvoltage}}
\subfigure[$\tau$ varied, $U_\mathrm{LED} = 7.6\mathrm{V}$]{\includegraphics[width=0.32\textwidth,angle=-90]{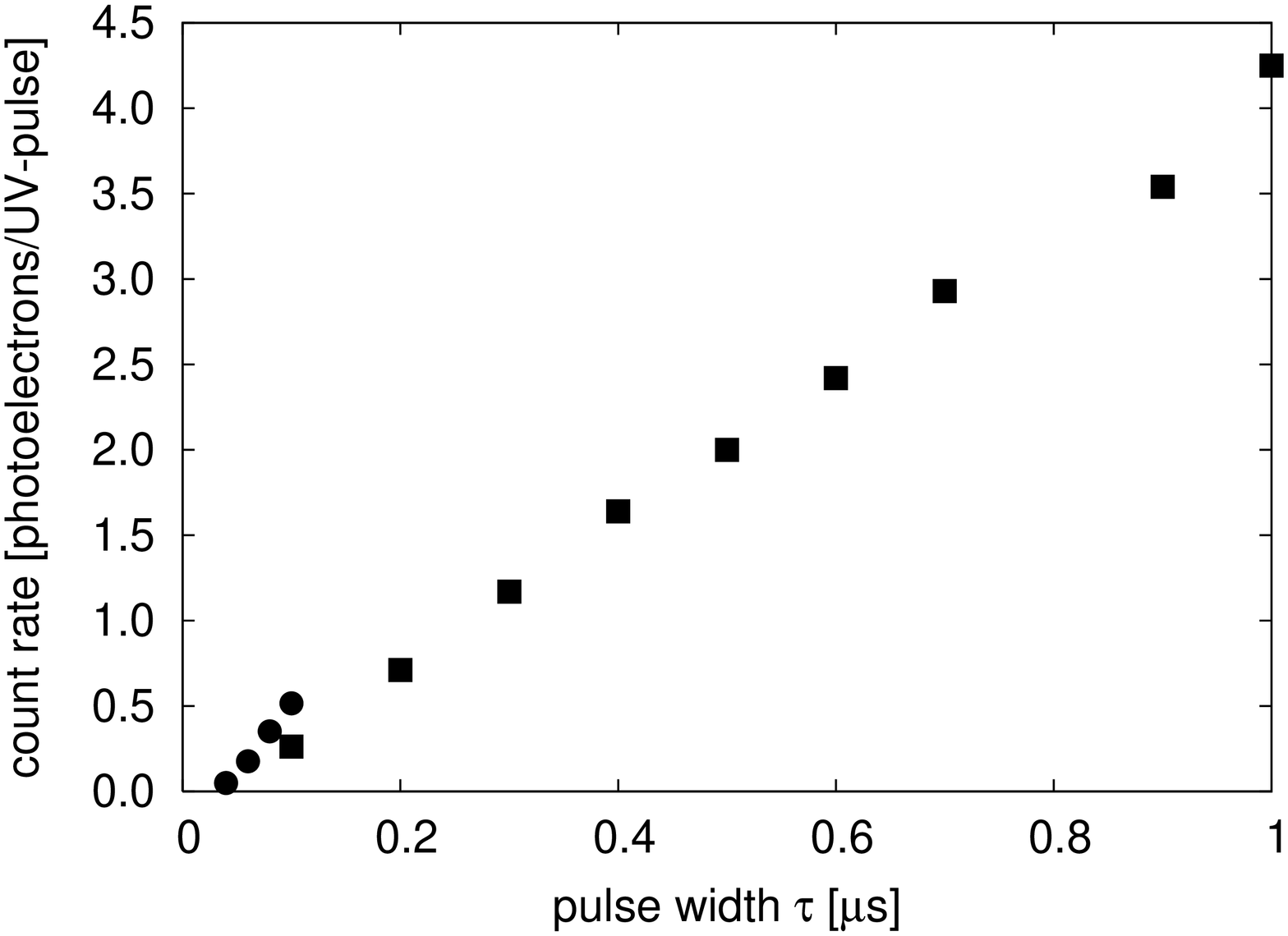}\label{fig:ledpulsewidth}}
\caption{Number of photoelectrons as a function of driving voltage
  $U_\mathrm{LED}$ supplied to the UV LED  (a) and as a function of the pulse
  width $\tau$ (b), both at a pulse repetition rate of $\unit[1]{kHz}$ and $E_\mathrm{surplus}=\unit[0.08]{keV}$. At $\tau = \unit[0.1]{\mu s}$ the alignment of the UV LED was changed, as indicated by the change of symbols (circles and squares, respectively) in figure \ref{fig:ledpulsewidth}.}
\end{figure}

 Figure \ref{fig:multiplepeaks} shows an example of the energy distribution of
 photoelectrons reaching the detector when the {\Mac} is in
 transmission, {\ie}, its threshold is lower than the kinetic energy of the
 photoelectrons ($qU_\mathrm{spec} < qU_\mathrm{source}$). The pulse width of the UV LED
 was $\tau = \unit[200]{ns}$ at a voltage of $U_\mathrm{LED}=\unit[7.6]{V}$. For $n$ coinciding
 photoelectrons, an energy equivalent to $E^n_\mathrm{photoel.} = n \cdot
 qU_\mathrm{source}$ will be measured. In the depicted case electron
 multiplicities $n$ up to $n = 4$ can be distinguished. However, single
 electrons are dominating at these settings.
By increasing or decreasing the forward current of the LED the number of
produced photoelectrons as well as their multiplicity distribution can be varied
and sufficiently many electrons can be created for high statistics
measurements with the {\Mac} (figure~\ref{fig:ledvoltage})\footnote{The UV LED
  can be operated beyond its specification DC voltage when operating with a
  small duty cycle. In these measurements pulse repetition rates were typically
  $\unit[1]{kHz}$ with pulse lengths up to $\unit[40]{\mu s}$ corresponding
  to duty cycles of $\unit[4]{\%}$ or less.}.

\begin{figure}[htb]
\centering
\subfigure[Single electron count rate]{\includegraphics[width=0.32\textwidth,angle=-90]{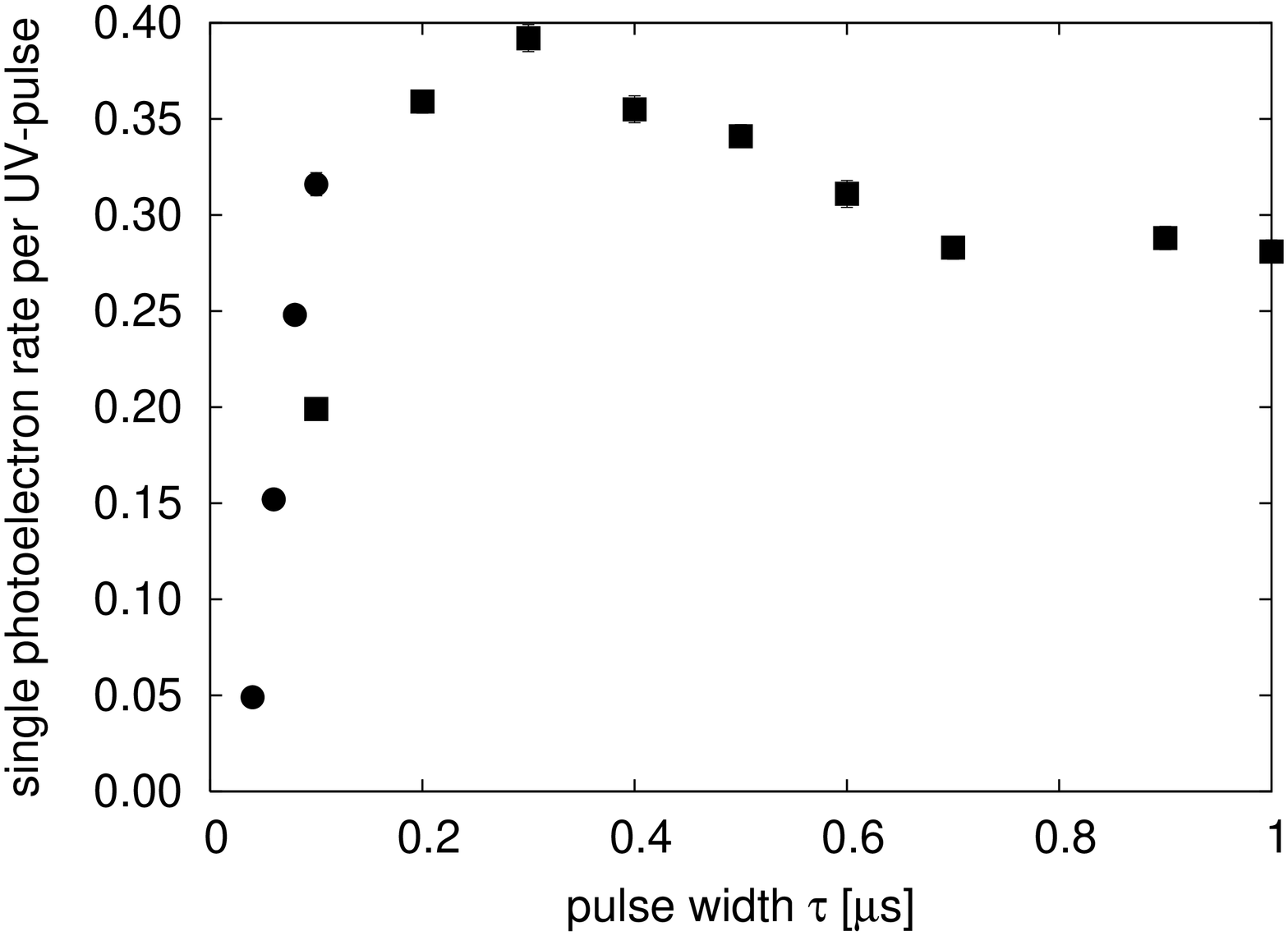}\label{fig:setotal}}
\subfigure[Single electron fraction]{\includegraphics[width=0.32\textwidth,angle=-90]{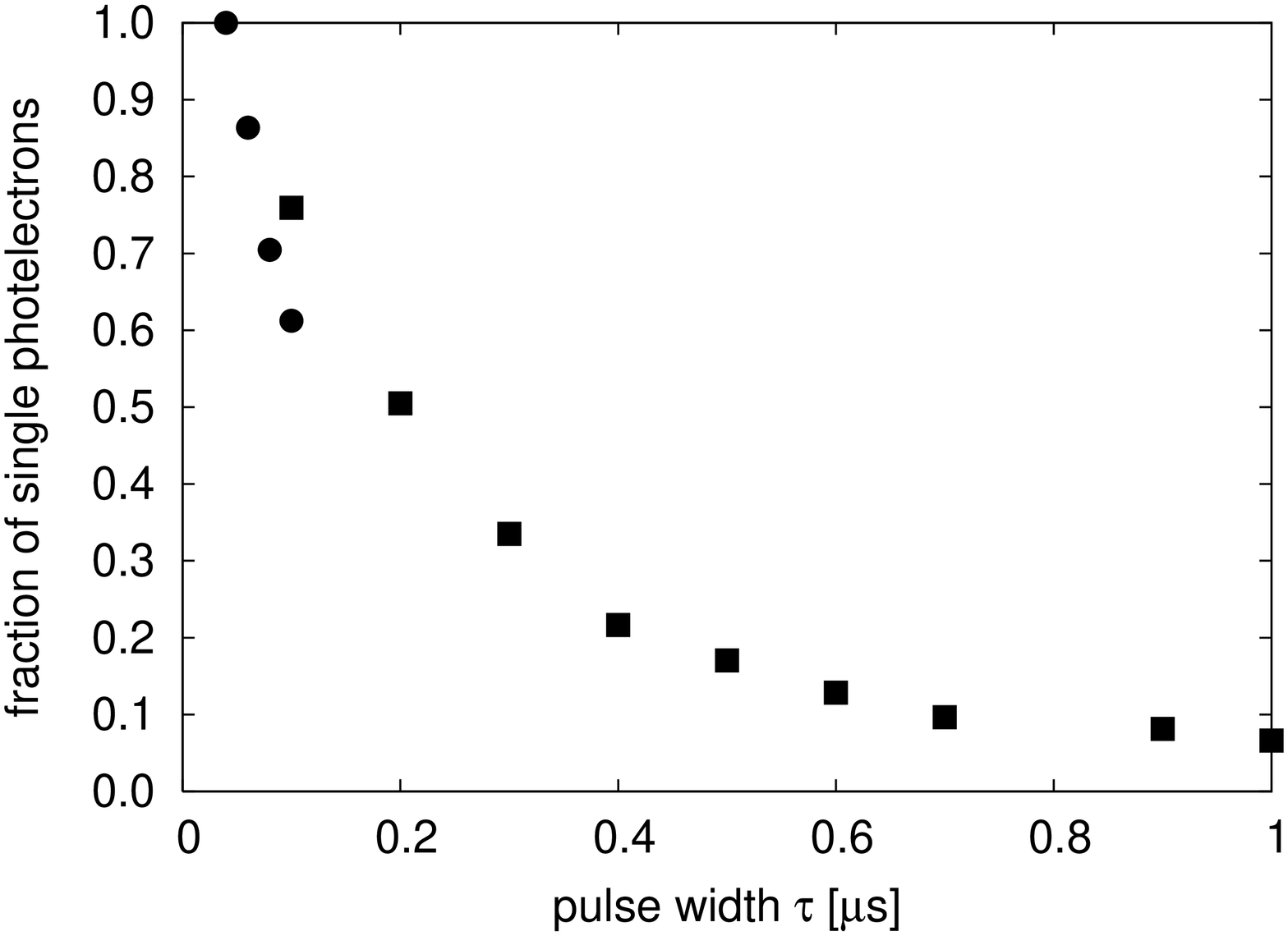}\label{fig:seratio}}
\caption{Total number (a) and relative fraction (b) of single photoelectrons per pulse
  as a function of pulse width $\tau$, measured with
  $U_\mathrm{LED}=\unit[7.6]{V}$. At $\tau = \unit[0.1]{\mu s}$ the alignment
  of the UV LED was changed, as indicated by the change of symbols (circles and squares, respectively) in both figures.}
\end{figure}

For advanced investigations of the transmission properties of MAC-E filters we
explored the possibility of time-of-flight (ToF) studies. For this purpose
single electrons with well defined energy and starting time are
needed. Figure~\ref{fig:ledpulsewidth} shows the average number of
photoelectrons created by the UV LED per pulse for short pulse widths at a
voltage of $U_\mathrm{LED} = \unit[7.6]{V}$. Even at a very short pulse width
of $\unit[40]{ns}$ photoelectrons are still emitted. The count rate for single
electrons only is shown in figure~\ref{fig:setotal}. For the small pulse widths
a plateau can be seen after an initial increase of the number of emitted
single electrons. Note that the number of single photoelectrons depends on the
minimal time difference needed to resolve to photoelectrons by our detector and data acquisition system, 
which in our case amounts to $\unit[755]{ns}$.

\begin{figure}[htb]
 \centering
\includegraphics[width=0.49\textwidth]{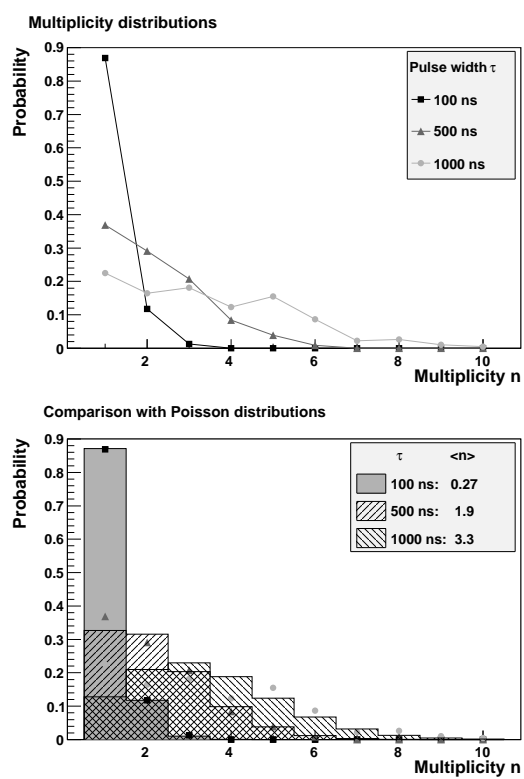}
\caption{Multiplicity distributions for selected pulse widths: probability for
  a measured event to contain $n$ electrons (top). The multiplicity
  distributions are roughly described by a Poisson distribution (bottom:
  symbols represent the measured data as in the figure at the top, Poisson distributions are described by histograms). $\langle n \rangle$ denotes the average multiplicity. The UV LED was operated with $U_\mathrm{LED}=\unit[7.6]{V}$.  \label{fig:multiplicity}}
\end{figure}

In figure~\ref{fig:seratio} it can be verified
that for the shortest pulse widths all emitted electrons come as single
electrons, albeit at a small rate. If higher electron multiplicities can be
tolerated then higher count rates of single electrons per pulse can be
achieved. The multiplicity distribution is shown in
figure~\ref{fig:multiplicity} (top) for pulse widths up to $\unit[1]{\mu
  s}$. The average multiplicity increases for longer pulse widths. The
multiplicity distributions are in reasonable agreement with Poisson statistics
(figure~\ref{fig:multiplicity}, bottom).


\section{Determination of the energy distribution and the angular emission 
of the photoelectrons at 18 keV}
\label{sec:tof-measurement}
\begin{figure}[!htb]
 \centering
\includegraphics[width=0.34\textwidth,angle=-90]{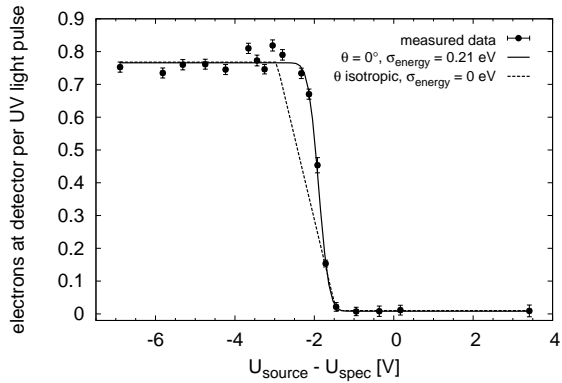}
\caption{Measured integral energy spectrum of photoelectrons. Also shown are two
  calculations for monoenergetic and isotropic photoelectron emission (dashed line)
  and for photoelectrons emitted along the normal of the photocathode but with
  a Gaussian energy spread of $\sigma_\mathrm{energy} = \unit[0.21]{eV}$ (solid line). The
  measurement is consistent with the latter.\label{fig:energyspectrum}}
\end{figure}

As mentioned above, a time-of-flight measurement requires the energy spread of the photoelectrons to be sufficiently
narrow. In order to verify this experimentally the photoelectron spectrum was
measured using the {\Mac}. The 265 nm LED was used for this measurement. By varying the retardation potential $U_\mathrm{spec}$ at
fixed photocathode potential $U_\mathrm{source}$, several spectra of the photoelectron
energy were recorded. For each measurement point the detected photoelectron
count rate was determined by weighting the counts of each electron peak in the
energy spectrum according to its multiplicity $n$. Figure
\ref{fig:energyspectrum} shows the total photoelectron count rate as a
function of the surplus energy $E_\mathrm{surplus} = q\,(U_\mathrm{source} -
U_\mathrm{spec})$ above the filter threshold. The width of this spectrum is
compared with the nominal energy resolution of $\Delta E = \unit[1.5]{eV}$ of 
the {\Mac}, which is defined for a monoenergetic electron source filling the full forward solid 
angle. 
The experimentally observed width may differ due to two main contributions:
\begin{itemize}
\item The observed width is smaller, since the electrons do not fill the full forward solid angle:
  If the angular distribution at the point of highest magnetic field
  $B_\mathrm{max}$, \ie, in the first
  superconducting magnet, would fill the full forward solid angle  ({\it e.g.} for an isotropically emitting radioactive source),
  then the residual cyclotron energy in the analyzing plane (and thus the non-analyzable fraction of the energy) would be $\Delta E = \unit[1.5]{eV}$,
  which is defined as the nominal resolution of the {\Mac}.
However, in our case, we do not expect that the angular distribution in the high magnetic 
field is filling the full forward solid angle
since the photoelectrons emitted at the photocathode are
immediately accelerated by the applied potential of $\unit[18]{kV}$ along the
E-field axis, and the B-field is nearly collinear with the E-field. Anyway, in the adiabatic case the cyclotron motion of the electron
around a magnetic field line averages out any transverse electrical field. Therefore, no transverse energy 
is picked up and the maximum angle is of order
$\theta_\mathrm{max} \approx \arctan \left( \sqrt{\frac{\unit[0.5]{eV}}{\unit[18]{keV}}} \right)=
  \unit[0.3]{^\circ}$, assuming a maximum initial transversal energy of order
  $\unit[0.5]{eV}$. Due to the transformation of parallel into transverse energy 
  according to equation~(\ref{equ:magn-moment}) when
  going from $B_\mathrm{start} \approx \unit[0.02]{T}$ at the location of the
  photoelectron source to the high field of $B_\mathrm{max} = \unit[6]{T}$
  in the magnet this angle corresponds to a maximum angle in the high field of
  $\approx \unit[5]{^\circ}$, which is far from the full forward solid angle case.

\item The observed width may be even larger than expected from the filled forward solid angle due to an energy spread of the electrons. In our
case, the photoelectrons obtain different energies due to the finite width of the photon energy ($\mathrm{FWHM} =
\unit[0.27]{eV}$) and the local inhomogeneity of the work function of the photocathode,
 A further broadening of the measured energy width may originate from potential energy losses of the electrons inside the photocathode and the ripple of the high-voltage power 
  supplies (${\mathcal O} (\unit[100]{mV})$).
\end{itemize}

These two contributions to the width will result in different shapes of the
energy spectrum. Calculations of the energy spectrum for two cases are also
shown in figure~\ref{fig:energyspectrum}:  The dashed line illustrates
the transmission function of a {\Mac} for 
an isotropically emitting but monoenergetic electron source. The full
line shows the expected transmission function for a single angle
electron source with emission along the normal of
the photocathode ($\theta = \unit[0]{^\circ}$) having an initial energy
spread of the photoelectrons described by a Gaussian broadening with $\sigma_\mathrm{energy}
= \unit[0.21]{eV}$. The latter curve fits much better to the data.

Although the small number of our data points and their limited statistics do not allow to really deconvolve the energy distribution of our 
photoelectron source, we can state that our data are well described by assuming that the photoelectrons are created with an energy spread 
of just $\sigma_\mathrm{energy} = \unit[0.21]{eV}$
and that their momentum vector is nearly parallel to the magnetic field lines after acceleration. Fitting these assumptions 
(Gaussian energy distribution and no angular spread) to the data yields a fit error for the Gaussian spread of $\Delta \sigma_\mathrm{energy} = \unit[\pm 0.02]{eV}$

This small energy spread and narrow angular distribution of the
photoelectrons, together with pulse lengths shorter
than $\unit[100]{ns}$ and a fast data acquisition system with a time
resolution of $\Delta t_\mathrm{DAQ,\, analysis} \approx \unit[60]{ns}$ for single electrons, allows us to perform a high-resolution determination of the
time of flight for photoelectrons passing the {\Mac} at different
surplus energies. Expected time-of-flight values range between $\approx \unit[200]{ns}$ for surplus
energies of $\unit[500]{eV}$ to several microseconds at few eV.

To obtain this time-of-flight spectrum we analyzed the data taken for the scan
of the energy spectrum of the photoelectrons with respect to the arrival
time distributions of the electrons.
The start signal for the measurement of arrival times was provided by the
trigger output of the function generator powering the UV LED. This control
signal was recorded on the second channel of the Flash-ADC card. The stop time
was given by the time of an electron signal from the detector preamplifier
recorded in the first channel of the Flash-ADC.

\begin{figure}[!htb]
 \centering
\subfigure[]{\includegraphics[height=0.49\textwidth,angle=-90]{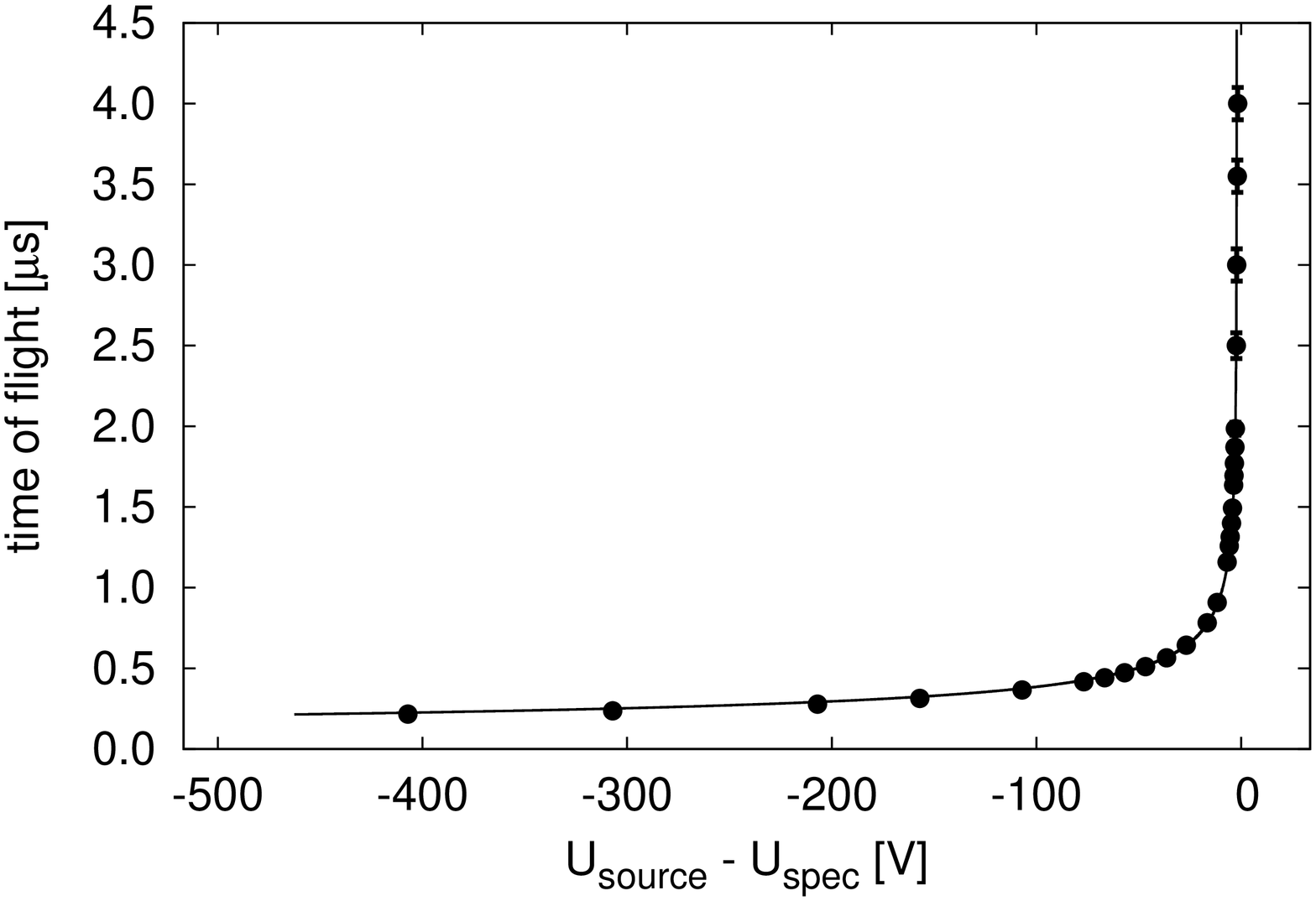}\label{fig:tof-measurement-a}}
\hfill
\subfigure[]{\includegraphics[height=0.49\textwidth,angle=-90]{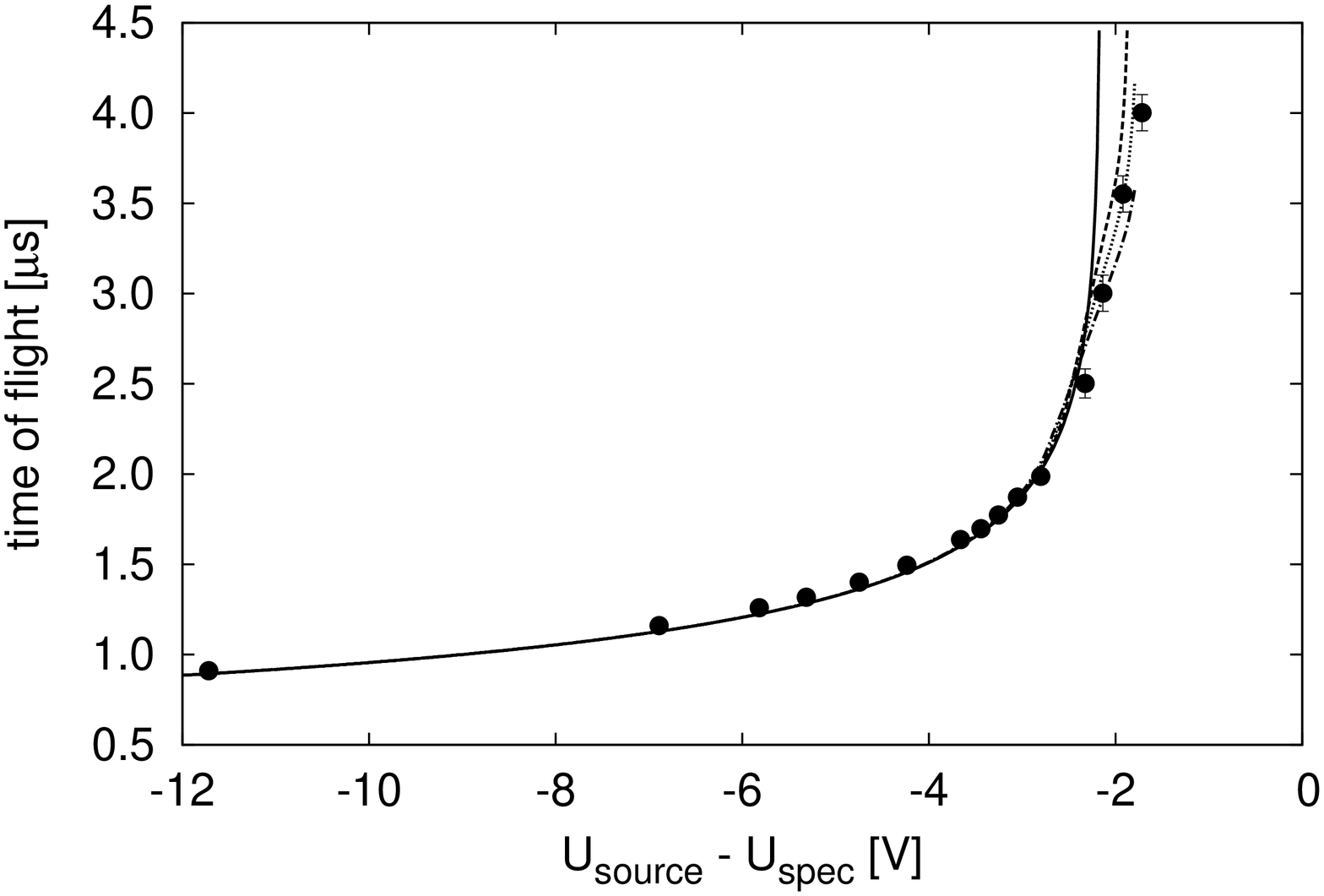}\label{fig:tof-measurement-b}}
\caption{Mean time of flight of photoelectrons passing the electrostatic filter
  versus surplus energy with respect to the filter potential
  ($E_\mathrm{surplus} = q(U_\mathrm{source} - U_\mathrm{spec})$) with $\tau =
  \unit[200]{ns}$ and $U_\mathrm{LED} = \unit[7.6]{V}$. (a)
  Comparison between measured values ($\bullet$) and values calculated
  according to equation~(\ref{equ:tof-integration}) (solid
  line) over a wide range of surplus energies. (b) For large time-of-flight
  values, corrections due to an energy smearing of the photoelectrons become
  relevant. Calculations for varying widths of a Gaussian broadening of the
  surplus energy are shown: $\sigma_\mathrm{energy} = \unit[0.15]{eV}$ (dashed
  line), $\unit[0.20]{eV}$ (dotted line), $\unit[0.25]{eV}$ (dash-dotted
  line), and no broadening (solid line). A good agreement between measurement and
  calculation is achieved for $\sigma_\mathrm{energy} = \unit[0.20]{eV}$.}
\label{fig:tof-measurement}
\end{figure}

Figure \ref{fig:tof-measurement} presents the time-of-flight spectrum for
photoelectrons over a wide range of surplus energies above the filter
potential using the mean time of flight for each surplus energy.
Electrons with large surplus energies $qU_\mathrm{source} -
qU_\mathrm{spec} = E_\mathrm{surplus} \approx 410~\mathrm{eV}$ reach the
detector about $\unit[220]{ns}$ after the UV light pulse irradiated the
photocathode, whereas electrons with small energies of $E_\mathrm{surplus}
\approx 1.7~\mathrm{eV}$ arrive after about $\unit[4]{\mu s}$. The
measurement is described well by a calculation based on

\begin{equation}
t_\mathrm{stop} - t_\mathrm{start} = \int_{x_\mathrm{start}}^{x_\mathrm{stop}} \frac{\mathrm{d}x}{v(x)} =
 \int_{x_\mathrm{start}}^{x_\mathrm{stop}} \sqrt{\frac{m_e}{2q(U_\mathrm{source}-U(x))}} ~~ \mathrm{d}x.
 \label{equ:tof-integration}
\end{equation}
Here we assumed that the velocity of the electron is parallel to the
magnetic field lines and thus defined by the gain of the 
kinetic energy $E_\mathrm{kin}$ in the electric potential $U$. This
corresponds to the expectation discussed above
that the photoelectrons are emitted with a negligible amount of transversal energy from 
the photocathode and accelerated by an
electric field parallel to the magnetic field lines.

Due to the broadening of the initial photoelectron energy distribution this calculation
deviates from the measurement for small excess energies. When taking the
broadening into account by folding equation~(\ref{equ:tof-integration}) with a
Gaussian energy distribution the calculation also agrees with the measurement
at small excess energies (figure~\ref{fig:tof-measurement-b}). The best
agreement is achieved
for a Gaussian width of $\sigma_\mathrm{energy} = \unit[0.2]{eV}$, which matches well the Gaussian width extracted from the integral energy spectrum.

\begin{figure}[!htb]
 \centering
\includegraphics[height=0.49\textwidth,angle=-90]{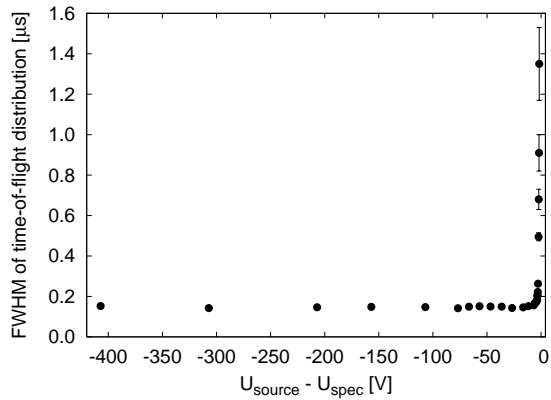}
\caption{Full width at half maximum (FWHM) of the measured time-of-flight distribution of photoelectrons at different
  surplus energies. For $U_\mathrm{source} - U_\mathrm{spec} \ll 0$ the FWHM of the measured time-of-flight distribution is limited by the full start time uncertainty of the electrons defined by the width of the UV light pulses ($\tau = \unit[200]{ns}$) and by the uncertainty of the arrival time determination related to the data acquisition system. \label{fig:width-arrivaltimes}}
\end{figure}

Figure~\ref{fig:width-arrivaltimes} shows the spread of the arrival time distributions for a given
surplus energy. The increase of the width of the distributions at small excess
energy again reflects the effect of the spread in initial photoelectron energy.

These measurements show that with a photoelectron source based on UV light from a modern LED irradiating a stainless steel cathode the time of flight of single electrons can be determined in \Mac s.

\section{Discussion of results and outlook}
\label{sec:outlook}
We have demonstrated that by using a deep-UV light emitting diode illuminating
a stainless steel cathode photoelectrons, which are well-defined
in energy and time, can be created with rates covering a large dynamical 
range. We accelerated these photoelectrons by an electric field to
a typical energy of $\unit[18]{keV}$. 

We determined the energy distribution of the photoelectrons in two ways, with both methods 
yielding consistent Gaussian widths of about $\sigma_\mathrm{energy} = \unit[0.2]{eV}$. 
This value may probably be further improved by using a cathode
material with a better defined work function ({\it e.g.} 
a monocrystalline surface instead of stainless steel in a better vacuum).
We achieved the time definition by electrically 
pulsing the UV LED. Pulse lengths from 40~$\mu$s  down to 40~ns were applied, 
resulting in a corresponding definition of the photoelectron start time. 
To reach high total photoelectron rates we used pulse repetition rates
of typically 1~kHz. By choosing the operating parameters of the UV LED 
like the operating voltage (and thus the forward current), pulse width 
and pulse repetition rate we obtained photoelectron rates from several
Hz up to several 10~kHz on average, with a maximum rate during the pulse
of $\unit[5]{electrons/\mu s}$. In particular, the operating parameters 
can be chosen such that only -- or at least predominantly -- 
single photoelectrons per pulse are achieved (figure~\ref{fig:seratio}), 
which allows to minimize the pile-up ratio and results in very good timing
({\eg}, for time-of-flight measurements). 

At the Mainz spectrometer (an electrostatic retardation spectrometer of MAC-E filter type) we measured
the transmitted electron rate as a function
of the retardation energy (integral {\Mac} mode), 
and the corresponding time-of-flight of the electrons (non-integral
MAC-E-TOF mode). We found that the observed integral 
as well as the time-of-flight spectrum agree well with
the expectations considering an initial Gaussian energy
width of the photoelectrons of $\sigma_\mathrm{energy} = \unit[0.2]{eV}$  
(compare figures~\ref{fig:energyspectrum} and \ref{fig:tof-measurement-b}).
These investigations were performed especially 
in view of an application as a test source for the large {\Mac} of 
the {\KE} (see Ref. \cite{disskv}). The achieved line width, the small angular emittance 
as well as the obtained time-of-flight precision match well the requirements to determine precisely 
the local electric retarding potential over the $\unit[64]{m^2}$ large analyzing plane of the KATRIN main spectrometer by
the onset of the transmission as well as to validate the correct evolution of the electric retarding potential along the electron trajectory by the time-of-flight method.

In a further development of this photoelectron source 
the additional feature of creating photoelectrons at well-defined trans\-versal energy can be accomplished by choosing a suitable configuration of
the electric and magnetic fields at the location of the photocathode. In the setup presented in this work magnetic and electric field were
essentially parallel at the photocathode, resulting in electrons with
negligible transversal energy. For different field configurations using a fast non-adiabatic acceleration in a strong electrical field transversal
energies $> 0$ can be realized. Such an electron source
allows to probe the transmission properties of a {\Mac} in much more detail (\eg, in dependence of the
transversal energy and the correctness of the adiabatic transformation according to equation~(\ref{equ:magn-moment})).
We will report on first experimental studies with a proof of principle in a forthcoming publication \cite{fiber-egun}. By combining fast pulsing with angular selectivity a powerful calibration tool for the KATRIN experiment and 
other applications may thus be achieved. 
\ack{This work was supported by the German Federal Ministry of Education
and Research under grant number 05CK5MA/0. The authors affiliated to WWU
M\"unster wish to thank the members of AG Quantum/Institut f\"ur Physik,
Johannes Gutenberg-Universit\"at Mainz, for their kind hospitality and for
giving us the opportunity to carry out these measurements in their laboratory.} 

\section*{References}
\begin{thebibliography}{10}
%
\bibitem{hsu-hirshfield} T. Hsu and J. L. Hirshfield, \href{http://link.aip.org/link/?RSINAK/47/236/1}{\emph{Electrostatic energy analyzer using a nonuniform axial magnetic field}}, Rev. Sci. Instrum. {\bf 47} (1976) 236
%
\bibitem{beamson}G. Beamson, H. Q. Porter and D. W. Turner, \href{http://dx.doi.org/10.1088/0022-3735/13/1/018}{\emph{The collimating and magnifying properties of a superconducting field photoelectron spectrometer}}, J. Phys. {\bf E\,13} (1980) 64; 
corrigendum: J. Phys. {\bf E\,14} (1981) 256;\newline
G. Beamson, H. Q. Porter and D. W. Turner, \href{http://www.nature.com/nature/journal/v290/n5807/pdf/290556a0.pdf}{\emph{Photoelectron spectromicroscopy}}, Nature {\bf 290} (1981) 556
%
\bibitem{kruit-read} P. Kruit and F. H. Read, \emph{Magnetic field paralleliser for $2\pi$ electron-spectrometer and electron-image magnifier}, J. Phys. {\bf E\,16} (1983) 313
%
\bibitem{picard-nimb} A. Picard \etal, \emph{A solenoid retarding spectrometer with high resolution and transmission for keV electrons}, Nucl. Instr. Meth. {\bf B\,63} (1992) 345
%
\bibitem{lobashev85} V. M. Lobashev and P. E. Spivak, \href{http://dx.doi.org/10.1016/0168-9002(85)90640-0}{\emph{A method for measuring the electron antineutrino rest mass}}, Nucl. Instr. Meth. {\bf A\,240} (1985) 305
%
\bibitem{kraus} Ch. Kraus, B. Bornschein, L. Bornschein, J. Bonn, B. Flatt, A. Koval\'{i}k, B. Ostrick, E. W. Otten, J. P. Schall, Th. Th\"ummler, Ch. Weinheimer, \href{http://www.springerlink.com/content/x73874485323j14m/fulltext.pdf}{\emph{Final results from phase II of the Mainz neutrino mass search in tritium $\beta$ decay}}, Eur. Phys. J. {\bf C\,40} (2005) 447
%
\bibitem{lobashev03} V. M. Lobashev, \href{http://dx.doi.org/10.1016/S0375-9474(03)00985-0}{\emph{The search for the neutrino mass by direct method in the tritium beta-decay and perspectives of study it in the project KATRIN}}, Nucl. Phys. {\bf A\,719} (2003) C153
%
\bibitem{kdr}The KATRIN collaboration (J. Angrik \etal), \emph{KATRIN Design Report 2004}, FZKA Scientific Report {\bf 7090} (2005), URL: \url{http://bibliothek.fzk.de/zb/berichte/FZKA7090.pdf}
%
\bibitem{otten-weinh} E. W. Otten and C. Weinheimer, \href{http://dx.doi.org/10.1088/0034-4885/71/8/086201}{\emph{Neutrino mass limit from tritium beta decay}}, Rep. Prog. Phys. {\bf 71} (2008) 086201
%
\bibitem{picard-zphys} A. Picard \etal, \emph{Precision measurement of the conversion electron spectrum of ~{$^{83m}Kr$} with a solenoid retarding spectrometer}, Z. Phys. {\bf A\,342} (1992) 71
%
\bibitem{campbell} J. L. Campbell and T. Papp, \emph{Widths of the atomic K--N7 levels}, Atomic Data and Nuclear Data Tables {\bf 77} (2001) 1
%
\bibitem{diss_bea}B. Ostrick, \emph{Eine kondensierte $^{83m}$Kr-Kalibrationsquelle f\"ur KATRIN},
Ph.D. thesis, Westf\"alische Wilhelms-Universit\"at M\"unster, 2009, and to be published
%
\bibitem{bonn} J. Bonn, L. Bornschein, B. Degen, E. W. Otten, Ch. Weinheimer, \emph{A high resolution electrostatic time-of-flight spectrometer with adiabatic magnetic collimation}, Nucl. Instr. Meth. {\bf A\,421} (1999) 256
%
%
\bibitem{deng} J. Deng \etal, \href{http://jjap.ipap.jp/link?JJAP/46/L263/}{\emph{247 nm Ultra-Violet Light Emitting Diodes}}, Jpn. J. Appl. Phys. {\bf 46} (2007) L263
%
\bibitem{gaska-zhang-2005} R. Gaska and J. P. Zhang, \href{http://dx.doi.org/10.1117/12.638266}{\emph{Deep-UV LEDs: Physics, Performance and Applications}}, Device and Process Technologies for Microelectronics, MEMS, and Photonics IV, Proceedings of SPIE Vol. {\bf 6037} (2005) 603706
%
\bibitem{shatalov-zhang-gaska-2003} M. Shatalov \etal, \href{http://link.aip.org/link/?APPLAB/82/167/1}{\emph{Time-resolved electroluminescence of AlGaN-based light-emitting diodes with emission at 285 nm}}, Appl. Phys. Lett. {\bf 82} (2003) 167
%
\bibitem{seoul-led-265} Seoul Semiconductor Co., Ltd., \emph{Specification document for UV LED model series T9B26*}, 2006, URL: \url{http://www.socled.com}
%
\bibitem{seoul-led-255} Seoul Semiconductor Co., Ltd., \emph{Specification document for UV LED model series T9B25*}, 2006, URL: \url{http://www.socled.com}
%
\bibitem{wei92} 
Ch. Weinheimer, M. Schrader, J. Bonn, Th. Loeken and H. Backe,
\emph{Measurement of energy resolution and dead layer thickness of $LN_2$-cooled PIN photodiodes}, Nucl. Instr. Meth. {\bf A\,311} (1992) 273

\bibitem{flatt-paper} B. Flatt, \emph{Voruntersuchungen zu den Spektrometern des KATRIN-Experiments}, Ph.D. thesis, Johannes-Gutenberg Universit\"at
  Mainz, 2004
%
\bibitem{disskv} K. Valerius, \emph{Spectrometer-related background processes and their suppression in the KATRIN experiment}, Ph.D. thesis, Westf\"alische Wilhelms-Universit\"at M\"unster, 2009
%
\bibitem{fiber-egun} K. Valerius \etal, to be published
%
\end{thebibliography}
\end{document}